\documentclass[a4paper]{jpconf}
\usepackage{wrapfig}

\usepackage{amsfonts,amssymb,amscd,amsmath}
\usepackage{amssymb}
\usepackage{amsmath}
\usepackage{graphicx}
\usepackage{graphics}
\usepackage{pifont}
\usepackage{hyperref}

\usepackage{adjustbox}

\def\le{\left(}
\def\ri{\right)}

\def\no{\nonumber}
\def\G{\Gamma}
\def\rar{\rightarrow}

\def\e{\epsilon}
\def\f12{\frac{1}{2}}

\def\pd{\partial}
\def\ve{\varepsilon}

\def\L{\lambda}

\usepackage{graphicx}
\begin{document}
\title{Alternative method of Reduction of the Feynman Diagrams to a set of Master Integrals}

\author{Julio Borja and Igor Kondrashuk}

\address{Grupo de F\'isica de Altas Energ\'ias, Departamento de Ciencias B\'asicas,  \\ 
         Universidad del B\'\i o-B\'\i o,  Campus Fernando May, 
          Casilla 447, Chill\'an, Chile}

\ead{igor.kondrashuk@gmail.com}

\begin{abstract}
We propose a new set of Master Integrals which can be used as a basis for certain multiloop calculations in massless gauge field theories. 
In these theories we consider three-point Feynman diagrams with arbitrary number of loops. The corresponding multiloop integrals may be decomposed in terms of this set of the Master Integrals. 
We construct a new reduction procedure which we apply to perform this decomposition.
\end{abstract}

\section{Introduction}

We describe the idea of an alternative method to calculate three-point vertex in a massless gauge field theory. At the first step, the method includes an algorithm to 
perform Lorentz algebra in the position space. This algorithm may be applied to the double ghost vertex of ${\cal N} = 4$ SYM in Landau gauge. The result of such an application will be a decomposition 
of this vertex in terms of basis elements. We describe these basis elements. The proposed algorithm is simple for programming.  At the next step, we show how  the proposed basis elements 
may be represented in terms of the integrals corresponding to the triangle ladder diagrams in $d=4-2\varepsilon$ dimensions.

\begin{figure}[ht!]  
\centering\includegraphics[scale=0.55]{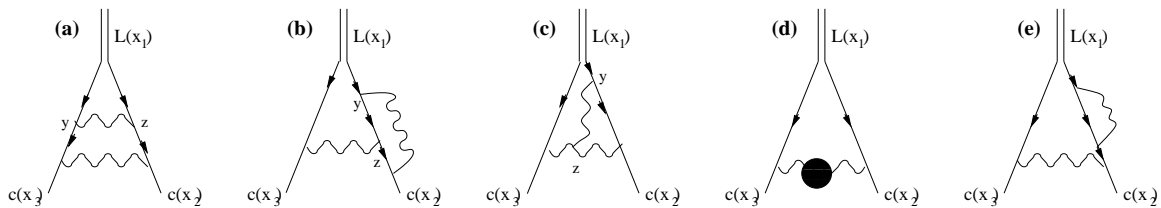}
\caption{\footnotesize  Two-loop planar corrections to the double ghost vertex. Wiggly lines correspond to gluon propagators, straight lines correspond to ghost propagators.  
The black disc in (d) is for one-loop contribution in the renormalization of the vector propagator from scalar, spinor and ghost fields.}
\label{figure-1}
\end{figure}

In Refs. \cite{1,2,3,4} the calculation of the two-loop correction to this auxiliary double ghost vertex for ${\cal N} = 4$ super-Yang-Mills (SYM) theory has been done in the limit $\varepsilon \rar 0.$
The contributing diagrams are shown  in Fig.\ref{figure-1}. 
This vertex is finite in ${\cal N} =4$ SYM in the Landau gauge at any loop order, it does not have poles in $\varepsilon$ in    $d=4-2\varepsilon$ dimensions. The superficial divergence disappears  
due to property of the transversality of the gluon propagator in the Landau gauge and divergences in subgraphs disappear due to  ${\cal N} =4$  supersymmetry.

\begin{figure}[t!] 
\centering\includegraphics[scale=0.8]{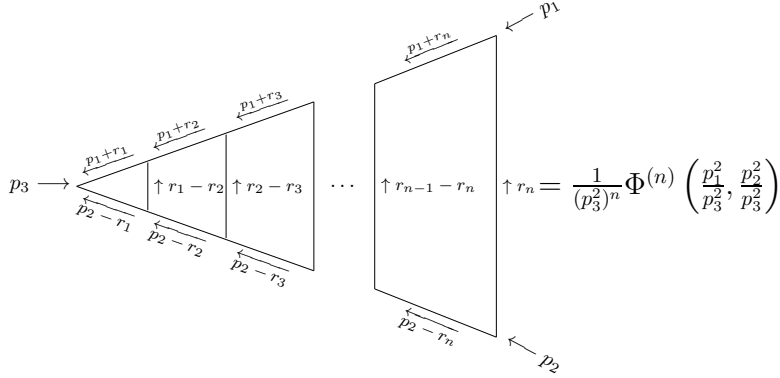}
\caption{\footnotesize  Triangle ladder diagrams for an arbitrary number $n$ of loops in the momentum space. Each line corresponds to certain power of scalar propagator 
in the momentum space $1/(p^2)^{\alpha}$, $\alpha$ is the index of the line. If the index is not written, it is equal to 1. Vertex has no structure.}
\label{figure-2}
\end{figure}

We show that all these contributions may be decomposed in terms of the integrals corresponding to the triangle ladder diagrams even in $d=4-2\varepsilon$ dimensions. 
This may help to analyse the planar limit of ${\cal N} = 4$ super-Yang-Mills theory. 
The triangle ladder diagrams have been studied  in $d=4$ dimensions in Refs. \cite{5,6,7} in the momentum space.  These triangle ladders with an arbitrary number of loops $n$ are shown in Fig. \ref{figure-2}
which is reproduced from Ref. \cite{8} with the same notation. 

\section{Mellin-Barnes  transforms of the momentum integrals of the triangle ladders}

It has been shown in Refs. \cite{9,10} at the diagrammatic level that the functions $\Phi^{(n)}\le x,y \ri$ obtained in Refs. \cite{6,7} for $d=4$, 
\begin{eqnarray*}
\Phi^{(n)}\le x,y\ri = -\frac{1}{n!\L}\sum_{j=n}^{2n}\frac{(-1)^j j!\ln^{2n-j}{(y/x)}}{(j-n)!(2n-j)!}\left[{\rm Li_j}
\le-\frac{1}{\rho x} \ri - {\rm Li}_j(-\rho y)\right], \\
\rho = \frac{2}{1-x-y+\L}, ~~~~ \L = \sqrt{(1-x-y)^2-4xy},
\end{eqnarray*}
of two variables that appear on the right hand side of Fig.\ref{figure-2} possess 
the property of invariance with respect to Fourier transformation. In Ref. \cite{11} the property of such an invariance of these functions has been formulated as 
\begin{eqnarray} \label{FI}
\frac{1}{[31]^2} \Phi^{(n)}\le \frac{[12]}{[31]},\frac{[23]}{[31]}\ri = \frac{1}{(2\pi)^4}\int~d^4p_1d^4p_2d^4p_3 ~ \delta(p_1 + p_2 + p_3) \times\no\\
\times e^{ip_2x_2} e^{ip_1x_1} e^{ip_3x_3} \frac{1}{(p_2^2)^2} \Phi^{(n)}\le \frac{p_1^2}{p_2^2},\frac{p_3^2}{p_2^2}\ri,  
\end{eqnarray}
where the notation  notation $[12] = (x_1-x_2)^2$ of Ref. \cite{2} is used. This property has been proven  via Mellin-Barnes transformation, that is, 
\begin{eqnarray*}
\Phi^{(n)}\le x,y\ri = \oint dz_2dz_3 x^{z_2} y^{z_3} {\cal M}^{(n)}\le z_2,z_3\ri, \Rightarrow \frac{1}{[31]^2} \Phi^{(n)}\le \frac{[12]}{[31]},\frac{[23]}{[31]}\ri = \\
\frac{1}{(2\pi)^8}\int~d^4p_1d^4p_2d^4p_3d^4x_5 e^{ip_2(x_2-x_5)} e^{ip_1(x_1-x_5)} e^{ip_3(x_3-x_5)}\frac{1}{(p_2^2)^2} \Phi^{(n)}\le \frac{p_1^2}{p_2^2},\frac{p_3^2}{p_2^2}\ri = \\
\frac{1}{(2\pi)^8}\int d^4p_1d^4p_2d^4p_3d^4x_5\oint dz_2dz_3 \frac{e^{ip_2(x_2-x_5)} e^{ip_1(x_1-x_5)} e^{ip_3(x_3-x_5)}}{(p_2^2)^{2+z_2+z_3} (p_1^2)^{-z_2} (p_3^2)^{-z_3}} {\cal M}^{(n)}\le z_2,z_3\ri  =  \\
\end{eqnarray*}
\begin{eqnarray*}
= \frac{(4\pi)^6}{(2\pi)^8}\int d^4x_5 \oint dz_2dz_3 ~ \frac{\G(-z_2-z_3)}{\G(2+z_2+z_3)} \frac{\G(2+z_2)}{\G(-z_2)} \frac{\G(2+z_3)}{\G(-z_3)}\\ 
\times\frac{2^{2z_2+2z_3-2(2+z_2+z_3)}{\cal M}^{(n)}\le z_2,z_3\ri}{[25]^{-z_2-z_3}[15]^{2+z_2} [35]^{2+z_3} }  = 
\oint~dz_2dz_3 ~ \frac{{\cal M}^{(n)}\le z_2,z_3\ri}{[12]^{-z_3}[23]^{-z_2}[31]^{2+z_2+z_3}}.  
\end{eqnarray*}

Such a simple proof suggests that the Mellin-Barnes transformation may be a helpful trick to study ladder diagrams in a non-integer dimension too. The Mellin-Barnes transforms of the 
momentum integrals corresponding to the ladder diagrams have been studied in Refs. \cite{12,13} for the ladders in $d=4$ dimensions and in Refs. \cite{14,15} in  $d=4-2\varepsilon$ 
dimensions with indices $1-\varepsilon$ on the rungs of the ladders. These integral transforms have been studied with a help of the diagrammatic relations  of Ref. \cite{5}, 
\begin{figure}[t!] 
\centering\includegraphics[scale=1.0]{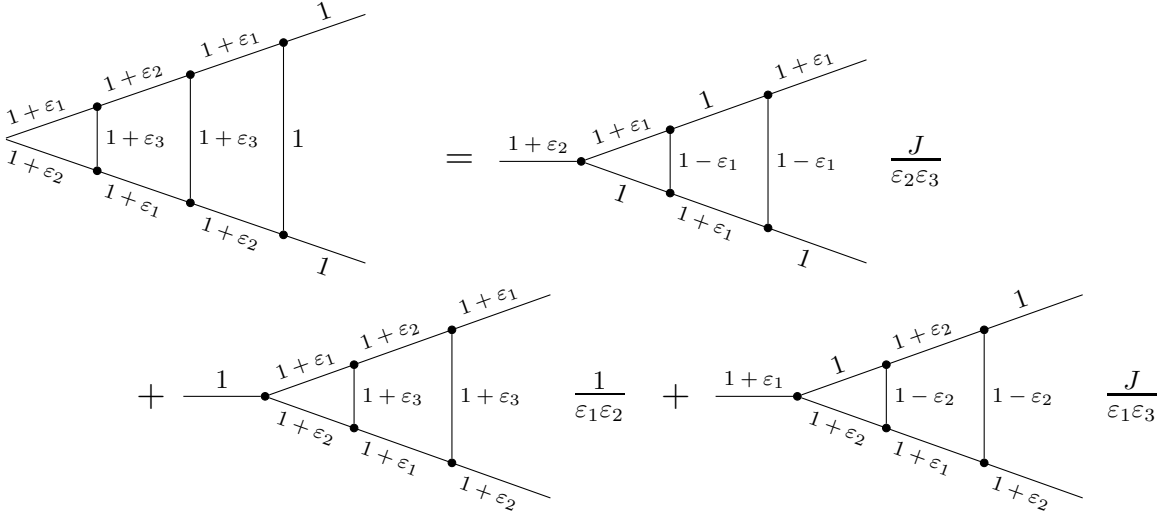}
\caption{\footnotesize  Loop reduction in $d=4$ dimensions.   Each line corresponds to certain power of scalar propagator in the position space $1/((y-z)^2)^{\alpha}$, $\alpha$ is the index 
of the line. The internal vertices $y_\mu$ and $z_\mu$ are integration points in the position space, they are depicted in boldface.   
The formula depicted is valid in $d=4$ for $\ve_1 + \ve_2 + \ve_3 = 0,$  with complex valued $\ve_1,\ve_2,\ve_3.$ Factor $J$ is defined in Eq.(\ref{def}). }
\label{figure-3}
\end{figure}
for the review of these relations one can see Ref. \cite{12}. This is a first known example when a three-point diagram with $n$ loops was represented as linear combination of diagrams with $n-1$ loops.   
As it has been shown in Ref. \cite{12}, the diagrammatic relation shown in Fig. \ref{figure-3} corresponds to the integral relation 
\begin{eqnarray} \label{ked}
\oint_C~dz_2dz_3~D^{(u,v)}[1+\ve_1-z_3,1+\ve_2-z_2,1+\ve_3] D^{(z_2,z_3)}[1+\ve_2,1+\ve_1,1+\ve_3] =  \no\\
J\left[ \frac{D^{(u,v-\ve_2)}[1-\ve_1]}{\ve_2\ve_3} +  \frac{D^{(u,v)}[1+\ve_3]}{\ve_1\ve_2}  + \frac{ D^{(u-\ve_1,v)}[1-\ve_2]}{\ve_1\ve_3}  \right],
\end{eqnarray}
where 
\begin{eqnarray*}
{D^{(z_2,z_3)}[\nu_1,\nu_2,\nu_3] = \frac{ \G \le -z_2 \ri \G \le -z_3 \ri \G \le -z_2 -\nu_2-\nu_3 + d/2 \ri \G \le -z_3-\nu_1-\nu_3 + d/2 \ri } {\G(d-\Sigma_i \nu_i)\Pi_{i} \G(\nu_i)  } }\times\no\\
 \times  \G \le z_2 + z_3  + \nu_3 \ri  \G \le  \Sigma \nu_i - d/2 + z_3 + z_2 \ri, 
\end{eqnarray*}
and where we have 
\begin{eqnarray} \label{def}
\ve_1 + \ve_2 + \ve_3 = 0, ~~~ D^{(u,v)}[1+\nu] \equiv  D^{(u,v)}[1,1,1+\nu],  ~~~ J =   \frac{\G(1-\ve_1)\G(1-\ve_2) \G(1-\ve_3) }{\G(1+\ve_1)\G(1+\ve_2) \G(1+\ve_3) }. 
\end{eqnarray}
Integral relation (\ref{ked}) has been proven via Barnes lemmas in \cite{16}. Going back from such a type of integral relations to diagrammatic relations we may construct 
the diagrammatic relations similar to the relation depicted in Fig. \ref{figure-3} in   $d=4-2\varepsilon$  dimensions. This observation makes the MB transformation to be a powerful trick 
to calculate ladder diagrams in $d=4-2\varepsilon$ exactly without expansion in Laurent series in $\ve$ \cite{16}.

\section{Algorithm}

The result for diagram $(c)$ of Fig.\ref{figure-1} may be found in Ref. \cite{4}. This diagram contains gluon self-interaction, for example, in $d=4$ the term 
\begin{eqnarray} \label{ked-2}
\frac{(31)_\nu}{[31]^2}\int~Dy~\frac{(2y)_\sigma}{[2y]^2}\frac{(1y)_\L}{[1y]^{2}}
\int~Dz \le\pd^{(z)}_\mu\Pi_{\rho\nu}(z3)\ri\Pi_{\rho\lambda}(zy) \Pi_{\mu\sigma}(z2), 
\end{eqnarray}
where the gluon propagator in $d = 4-2\ve$ in the position space is 
\begin{eqnarray*}
\Pi_{\rho\L}(zy)  =  \frac{g_{\rho\L}}{[yz]^{1-\e}} + 2(1-\e)\frac{(yz)_\rho (yz)_\L}{[yz]^{2-\e}}
\end{eqnarray*}
and the notation is $[yz] \equiv (y-z)^2, ~~~~ [y1] \equiv (y-x_1)^2,~~~ (yz)_\nu \equiv (y - z)_\nu,~~(31)_\nu \equiv (x_3-x_1)_\nu$ and $Dx \equiv \pi^{-\frac{d}{2}}d^d x$ 
is a measure of the integration in the position space. This measure is useful in massless theories \cite{2}.  The result for the diagram $(c)$ from Ref. \cite{4} in the limit $\varepsilon \rar 0$ is
\begin{eqnarray*}
\frac{13/2}{[12]^2[23]^2} + \frac{-33/2}{[12]^2[31]^2}  + \frac{15/2}{[23]^2[31]^2}  +  \frac{11}{[12][23][31]^2}  
+  \frac{-14}{[12][23]^2[31]} +  \frac{10}{[12]^2[23][31]}   + \left[\frac{6}{[12][23]^2} \right. \\
\left. + \frac{-6}{[12][31]^2}  +  \frac{[12]}{[23]^2[31]^2}  + \frac{4[23]}{[12]^2[31]^2} + \frac{-2}{[12]^2[31]}  + \frac{-5}{[23]^2[31]}   
+ \frac{-2[31]}{[12]^2[23]^2} \right] \frac{1}{[31]} \Phi^{(1)}\le \frac{[12]}{[31]},\frac{[23]}{[31]}\ri \\
+ \left[\frac{-2}{[12]^2[23]^2} +  \frac{6}{[12]^2[31]^2} +   \frac{4}{[23]^2[31]^2} +  \frac{1}{[12][23][31]^2} +  \frac{-1}{[12][23]^2[31]} 
+ \frac{-4}{[12]^2[23][31]} \right]\ln[12] \\
+ \left[\frac{-1/2}{[12]^2[23]^2} + \frac{-5}{[12]^2[31]^2}    +   \frac{-3/2}{[23]^2[31]^2}   +   \frac{-7/2}{[12][23][31]^2} +   \frac{2}{[12][23]^2[31]}   
+ \frac{7/2}{[12]^2[23][31]}  \right]\ln[23] \\
+ \left[\frac{5/2}{[12]^2[23]^2} + \frac{-1}{[12]^2[31]^2}    +   \frac{-5/2}{[23]^2[31]^2}   +   \frac{5/2}{[12][23][31]^2} +   \frac{-1}{[12][23]^2[31]}   
+ \frac{1/2}{[12]^2[23][31]}  \right]\ln[31].
\end{eqnarray*}
The gluon self-interaction term of Eq.(\ref{ked-2})  may be decomposed in a sum of  simpler terms which contain derivatives acting on the scalar propagator,  for example, 
\begin{eqnarray} \label{element}
\int~Dy~Dz~\frac{1}{[31]^{1-\epsilon}} \frac{1}{[2y]^{1-\epsilon}} \frac{1}{[3z]^{1-\epsilon}} \frac{1}{[z2]^{1-\epsilon}} 
\pd^{(y)}_{\mu}\frac{1}{[1y]^{1-\epsilon}} \pd^{(y)}_{\mu}\frac{1}{[yz]^{1-\epsilon}}.   
\end{eqnarray}
There are other terms with higher number of pairs of the contracted derivatives. We call the terms of such a type  basis elements.

Now we show how the basis elements may be decomposed in terms of the triangle ladder diagrams. The algorithm has already been described 
in Ref. \cite{10}. Here we describe the idea at level of formulas.  The trick of the reduction of the basis elements  to integrals corresponding to the triangle ladder diagrams 
may be shown for the element of Eq. (\ref{element})     
\begin{eqnarray} \label{element-2}
\int~Dy~Dz~\frac{1}{[31]^{1-\epsilon}} \frac{1}{[2y]^{1-\epsilon}} \frac{1}{[3z]^{1-\epsilon}} \frac{1}{[z2]^{1-\epsilon}} 
\pd^{(y)}_{\mu}\frac{1}{[1y]^{1-\epsilon}} \pd^{(y)}_{\mu}\frac{1}{[yz]^{1-\epsilon}}  = \no\\
\frac{1}{2}\int~Dy~Dz~\frac{1}{[31]^{1-\epsilon}} \frac{1}{[2y]^{1-\epsilon}} \frac{1}{[3z]^{1-\epsilon}} \frac{1}{[z2]^{1-\epsilon}} \times \no\\
\times \left[\pd_{(y)}^2\le\frac{1}{[1y]^{1-\epsilon}} \frac{1}{[yz]^{1-\epsilon}}\ri -  \le\pd_{(y)}^2\frac{1}{[1y]^{1-\epsilon}}\ri 
\frac{1}{[yz]^{1-\epsilon}} -  \frac{1}{[1y]^{1-\epsilon}} \le\pd_{(y)}^2\frac{1}{[yz]^{1-\epsilon}} \ri \right].
\end{eqnarray}
D'Alembert operator acting on the scalar propagator generates Dirac $\delta$-function. This means that the double integration disappears and the 
result reduces to the simple integration of three propagators in each of the terms on the right hand side of Eq. (\ref{element-2}). In more complicate 
cases the d'Alembertian at the end of these transformations may act on one of the external points of the triangle ladder diagram. As we have mentioned in the 
previous section, these integrals may be calculated exactly even in  $d=4-2\varepsilon$  dimensions via the MB transformation.

\section{Conjecture}

We make the conjecture that the diagrammatic relations in Fig. 1 of Ref. \cite{10} established in $d=4$ dimensions will maintain themselves in  $d=4-2\varepsilon$  
dimensions, too.  This may be proved via  Mellin-Barnes transformation with help of the trick presented in talk \cite{16}. On the other hand, the integrals on the right hand side 
of these relations are more easy to calculate than the ladder diagrams.  Such a calculation again may be done by using the MB transformation.  
The result of this chain of transformations may be considered as an independent crosscheck of the method described in Section 2. Thus, the algorithm we have 
proposed in Section 3 gives us the decomposition of the three-point double ghost vertex in terms of the ladder diagrams on which the d'Alembert operator acts as it is shown on Fig. 1 of Ref. \cite{10}, 
and these constructions may be calculated explicitly in  $d=4-2\varepsilon$  dimensions by two different methods. The first method is described in Section 2 and the
second method is described in this Section.

\section{Conclusion}

The advantage of this approach is that the proposed algorithm is easy to program. Algebraically, it reduces the integration to the integrals in the momentum space 
corresponding to the triangle ladder diagrams. We have explained in the previous Sections why such diagrams may be calculated explicitly via the Mellin-Barnes transformation
in  $d=4-2\varepsilon$ dimensions.  The method may serve in any massless field theory, not necessarily in ${\cal N} =4$ SYM only.

\ack{These results are based on the papers written in collaboration with Pedro Allendes (UdeC), Gorazd Cvetic (UTFSM), Ivan Gonzalez (UV), Bernd Kniehl (DESY), 
Anatoly Kotikov (JINR), Eduardo Notte-Cuello (ULS),  Ivan Parra-Ferrada (U de Talca), Marko Rojas-Medar (UBB), Ivan Schmidt (UTFSM), Alvaro Vergara (UBB). I.K. is very grateful to all of them.
This research is supported in part by Fondecyt (Chile) Grants Nos. 1040368, 1050512 and 1121030, by DIUBB (Chile) Grant Nos.  125009,  GI 153209/C  and GI 152606/VC.}

\end{document}